\documentclass[prd,superscriptaddress,amsfonts,amssymb,amsmath,showpacs]{revtex4-2}
\usepackage{bm}
\usepackage{amsfonts}
\usepackage{latexsym}
\usepackage[latin1]{inputenc}
\usepackage{graphicx}
\usepackage{amsmath}
\usepackage{palatino}
\usepackage{mathpazo}
\usepackage[british]{babel}
\usepackage{hhline}
\usepackage{multirow}
\usepackage{textcomp}
\usepackage{float}
\usepackage{booktabs}
\usepackage{dcolumn}
\usepackage{hhline}
\usepackage{multirow}
\usepackage{ragged2e}
\usepackage{hyperref}
\hypersetup{colorlinks,citecolor=blue}
\hypersetup{colorlinks=true,linkcolor=red,filecolor=magenta,    urlcolor=cyan}
\usepackage{amsmath}
\usepackage{xcolor}
\usepackage{orcidlink}
\usepackage{epsfig}
\usepackage{caption}
\usepackage{subcaption}
\usepackage{commath}
\def\jnl@style{\it}
\def\aaref@jnl#1{{\jnl@style#1}}

\def\aaref@jnl#1{{\jnl@style#1}}

\def\aj{\aaref@jnl{AJ}}                   
\def\apj{\aaref@jnl{ApJ}}                 
\def\apjl{\aaref@jnl{ApJ}}                
\def\apjs{\aaref@jnl{ApJS}}               
\def\apss{\aaref@jnl{Ap\&SS}}             
\def\aap{\aaref@jnl{A\&A}}                
\def\aapr{\aaref@jnl{A\&A~Rev.}}          
\def\aaps{\aaref@jnl{A\&AS}}              
\def\mnras{\aaref@jnl{Mon.~Not.~Roy.~Astron.~Soc.}}             
\def\prd{\aaref@jnl{Phys.~Rev.~D}}        
\def\prc{\aaref@jnl{Phys.~Rev.~C}}  
\def\prl{\aaref@jnl{Phys.~Rev.~Lett.}}    
\def\qjras{\aaref@jnl{QJRAS}}             
\def\skytel{\aaref@jnl{S\&T}}             
\def\ssr{\aaref@jnl{Space~Sci.~Rev.}}     
\def\zap{\aaref@jnl{ZAp}}                 
\def\nat{\aaref@jnl{Nature}}              
\def\aplett{\aaref@jnl{Astrophys.~Lett.}} 
\def\apspr{\aaref@jnl{Astrophys.~Space~Phys.~Res.}} 
\def\physrep{\aaref@jnl{Phys.~Rep.}}      
\def\physscr{\aaref@jnl{Phys.~Scr}}       
\def\commat{\aaref@jnl{Comm.~Math.~Phys.}}              
\def\science{\aaref@jnl{Science}}               
\def\cqg{\aaref@jnl{Classical Quant.~Grav.}}            
\def\jpcs{\aaref@jnl{JPCS}}                                     
\def\ijmpd{\aaref@jnl{Int.~J.~Mod.~Phys.~D}}                    
\def\grg{\aaref@jnl{Gen.~Relat.~Gravit.}}               
\def\rpp{\aaref@jnl{Rep.~Prog.~Phys.}}          
\def\npa{\aaref@jnl{Nucl.~Phys.~A}}        
\def\lrr{\aaref@jnl{Living Rev.~Rel.}}                   
\def\jcap{\aaref@jnl{J.~Cosmology Astropart.~Phys.}}    
\def\rmp{\aaref@jnl{Rev.~Mod.~Phys.}}   
\def\epjc{\aaref@jnl{Eur.~Phys.~J.~C}} 
\def\plb{\aaref@jnl{~Phy.~Lett.~B}} 
\def\mpla{\aaref@jnl{Mod.~Phy.~Lett.~A}} 
\def\arxiv{\aaref@jnl{arxiv.org}}

\allowdisplaybreaks[1]

\begin{document}
\color{black}       
\title{Noether Symmetry Approach in Scalar-Torsion \texorpdfstring{$f(T,\phi)$}{} Gravity }

\author{L.K. Duchaniya \orcidlink{0000-0001-6457-2225}}
\email{duchaniya98@gmail.com}
\affiliation{Department of Mathematics,
Birla Institute of Technology and Science-Pilani, Hyderabad Campus,
Hyderabad-500078, India.}

\author{B. Mishra\orcidlink{0000-0001-5527-3565}}
\email{bivu@hyderabad.bits-pilani.ac.in}
\affiliation{Department of Mathematics,
Birla Institute of Technology and Science-Pilani, Hyderabad Campus,
Hyderabad-500078, India.}

\author{Jackson Levi Said\orcidlink{0000-0002-7835-4365}}
\email{jackson.said@um.edu.mt}
\affiliation{Institute of Space Sciences and Astronomy, University of Malta, Malta, MSD 2080.}
\affiliation{Department of Physics, University of Malta, Malta.}

\begin{abstract}
The Noether Symmetry approach is applied to study an extended teleparallel $f(T,\phi)$ gravity that contains the torsion scalar $T$ and the scalar field $\phi$ in the context of an Friedmann-Lema\^{i}tre-Robertson-Walker space-time. We investigate the Noether symmetry approach in $f(T,\phi)$ gravity formalism with the specific form of $f(T,\phi)$ and analyze how to demonstrate a nontrivial Noether vector. The Noether symmetry method is a helpful resource for generating models and finding out the exact solution of the Lagrangian. In this article, we go through how the Noether symmetry approach enables us to define the form of the function $f(T,\phi)$ and obtain exact cosmological solutions. We also find the analytical cosmological solutions to the field equations, that is consistent with the Noether symmetry. Our results demonstrate that the obtained solutions enable an accelerated expansion of the Universe. We have also obtained the present value of the Hubble parameter, deceleration parameter, and effective equation of state parameter, which is fit in the range of current cosmological observations. 
 
\end{abstract}

\maketitle

\section{Introduction} \label{SEC-I}
General Relativity (GR) has gone through over a century of successfully describing the evolutionary processes of the Universe in the form of the $\Lambda$CDM model \cite{misner1973gravitation,Clifton:2011jh,Aghanim:2018eyx}, which is supported by overwhelming observational and fundamental precision tests. This scenario predicts a Universe that drives the big bang through an inflationary epoch and the well-known early Universe dynamics to eventually produce an accelerating late-time cosmology that is sourced by dark energy \cite{Riess:1998cb,Perlmutter:1998np}. $\Lambda$CDM describes dark energy through a cosmological constant $\Lambda$ which continues to have fundamental problems associated with it \cite{RevModPhys.61.1,Appleby:2018yci,Ishak:2018his} despite its observational successes. The next leading-order contribution to this late-time cosmology is cold dark matter (CDM), which primarily acts on galactic scales. Despite numerous decades-long efforts, this remains observational and undetected \cite{Baudis:2016qwx,Bertone:2004pz}. In the last few years, this has become all the more dire with a new challenge coming from the observational sector, which is the suggestion of tension in the value of the Hubble constant \cite{Bernal:2016gxb,DiValentino:2020zio,DiValentino:2021izs} as measured from local \cite{Riess:2019cxk,Wong:2019kwg}, early Universe sources \cite{DES:2017txv,Aghanim:2018eyx}. This continues to seemingly increase as an observational tension in the data \cite{Riess:2021jrx,Brout:2021mpj,Scolnic:2021amr}, and may permeate into other sectors of cosmology \cite{Abdalla:2022yfr,DiValentino:2020vvd}.

One possible way to confront this problem is to consider even further modifications to the matter sector, which would produce effective differences at particular epochs of the Universe, similar to inflation. However, another approach is to reconsider the concordance model description of gravity through modifications to GR \cite{Clifton:2011jh,Capozziello:2011et,CANTATA:2021ktz,Nojiri:2010wj}. Recently, considerable work has gone into a new setting in which to consider gravitational interactions, namely teleparallel gravity (TG). Here, the curvature associated with the Levi-Civita connection ($\mathring{\Gamma}^{\rho}_{\mu\nu}$, over-circles denote any quantities calculated with the Levi-Civita connection) is exchanged with the torsion produced by the teleparallel connection ($\Gamma^{\rho}_{\mu\nu}$) \cite{Aldrovandi:2013wha,Cai:2015emx,Krssak:2018ywd,Bahamonde:2021gfp}. This is a curvature-less connection that satisfies metricity. This means that all measures of curvature will turn out to be identically zero, such as the Ricci scalar $R(\Gamma^{\rho}_{\mu\nu}) = 0$. Saying that the regular Ricci scalar remains nonzero in general ($\mathring{R}(\mathring{\Gamma}^{\rho}_{\mu\nu}) \neq 0$). TG can be used with regular GR to produce a torsion scalar $T$, equal to the curvature-based Ricci scalar (up to a boundary term). Naturally, an action based on the torsion scalar will then be dynamically equivalent to GR, and it is thus called the \textit{Teleparallel Equivalent of General Relativity} (TEGR) since it produces the same dynamical equations as that of the Einstein-Hilbert action. 

Curvature-based modifications of GR have taken various forms over the years, with the most popular being $f(\mathring{R})$ gravity \cite{Sotiriou:2008rp,Faraoni:2008mf,Capozziello:2011et}. Similarly, TEGR can be directly generalized to $f(T)$ gravity \cite{Ferraro:2006jd,Ferraro:2008ey,Bengochea:2008gz,Linder:2010py,Chen:2010va,Bahamonde:2019zea,Duchaniya:2022rqu}. $f(T)$ gravity has the added advantage that it is a second-order theory in terms of the derivatives that appears in the equations of motion. In this context, it might also be interesting to add a scalar field $\phi$ to this general function since the resulting $f(T,\phi)$ Lagrangian will continue to be second order in these derivatives \cite{Geng:2011aj,Geng:2011ka,Otalora:2013dsa,Otalora:2013tba,PhysRevD.87.076006,Kadam:2022lgq,PhysRevD.97.104011}. This is the TG analog of $f(R,\phi)$ gravity \cite{Faraoni04} with the important distinction that here all equations of motion are second-order in nature. This setting of gravitational models has already been studied somewhat in works such as Refs.~\cite{Gonzalez-Espinoza:2021mwr,Gonzalez-Espinoza:2020jss,Gonzalez_Espinoza_2021ge,Gonzalez_Espinoza_2020}. However, the Noether symmetry considerations remain an open question for such classes of models. 

 In this work, we consider the Noether symmetry approach detailed in Refs.~\cite{PhysRevD.42.1091,PhysRevD.46.1391,Capozziello_1993,Capozziello_2000, Kucukakca_2013,Kadam_2023ns}. Through this approach, we will study potential cosmological evolution scenarios produced by particular models of this class of theories. Noether symmetries offer a tool to solve dynamical equations within cosmology, but more than that, it permits a way to produce models that have some motivation from the fundamental sector. This provides better motivation to study complex systems of equations of motion. Recently, Ref.~\cite{Dialektopoulos:2021ryi} studied the full classification of teleparallel Horndeski scalar-tensor theories of cosmology stemming from Refs.~\cite{Bahamonde:2019ipm,Bahamonde:2019shr}. This motivates us to analyze further this particular subclass of models in which a simpler form of the scalar field contribution is assumed.

We organize the work as follows; in Sec.~\ref{SEC-II}, we briefly discuss the technical details of TG and its formulation of $f(T,\phi)$ gravity, together with the formulation of the Friedmann equations for this setting. In Sec.~\ref{SEC-III}, we obtain the point-like Lagrangian and derive the Noether equations using the Euler-Lagrangian equations in configuration space $\mathcal{Q}=(a, T,\phi)$, leading to the cosmological equations of motion Eqs.~(\ref{8}--\ref{10}). In Sec.~\ref{SEC-IV}, we introduce the concept of Noether symmetries, leading to Sec.~\ref{SEC-V}, where these symmetries are studied for the present case. By determining the Noether vector for a specific form of $f(T,\phi)$, we also determine the exact solutions of the cosmological field equations in Sec.~\ref{SEC-VI}. Finally, we conclude with a summary of the main results in Sec.~\ref{SEC-VII}.

\section{Scalar-torsion \texorpdfstring{$f(T,\phi)$}{} gravity}\label{SEC-II}
Replacing the metric tensor as the fundamental variable with tetrad $e^{A}_{\,\,\,\,\mu}$ (inverse represented by $E_{A}^{\,\,\,\,\mu}$) fields together with a spin connection $\omega^{a}_{\,\,\, b\mu}$ as the dynamical variable, GR can be reformulated in the context of TG. The tetrad field $e^{A}_{\,\,\,\,\mu}$ (where Latin indices take on the values $A = 0, 1, 2, 3$ refer to coordinates on the tangent space) relates local Lorentz frames with the general spacetime manifold coordinates, which are denoted by Greek indices. The metric can then be built as
\begin{equation}
    g_{\mu \nu}=\eta_{AB} e^{A}_{\,\,\,\,\mu}~ e^{B}_{\,\,\,\,\nu}\,,
\end{equation}

where $\eta_{AB}$ represents the Minkowski metric. The tetrad must also meet the requirements of orthogonality ${\color{blue}E}_{A}^{\,\,\,\,\mu}  e^{B}_{\,\,\,\,\mu}=\delta_A^B$. Using the tetrad, the Levi-Civita connection can be substituted by the torsion-ful teleparallel connection, given by \cite{Krssak:2015oua}
\begin{equation}
    \Gamma^{\sigma}_{\,\,\,\,\nu\mu} := E_{A}^{\,\,\,\,\sigma}\left(\partial_{\mu}e^{A}_{\,\,\,\,\nu} + \omega^{A}_{\,\,\,\,B\mu} e^{B}_{\,\,\, \nu}\right)\,,
\end{equation}

where the spin connection acts to retain the local Lorentz invariance of the ensuing field equations, for a particular frame, called the Weitzenb\"{o}ck gauge, these components vanish identically. Using this connection, an analog of the Riemann tensor, which vanishes for the teleparallel connection, can be defined as an anti-symmetric operator on this connection through \cite{Hayashi:1979qx}
\begin{equation}
    T^{\sigma}_{\,\,\,\,\mu\nu} :=2 \Gamma^{\sigma}_{\,\,\,\,[\nu\mu]}\,.
\end{equation}
Using this torsion tensor, the torsion scalar can be defined as \cite{Krssak:2018ywd,Cai:2015emx,Aldrovandi:2013wha,Bahamonde:2021gfp}
\begin{equation}\label{eq:torsion_scalar_def}
    T:=\frac{1}{4} T^{\alpha}_{\,\,\,\,\mu\nu} T_{\alpha}^{\,\,\,\,\mu\nu} + \frac{1}{2} T^{\alpha}_{\,\,\,\,\mu\nu} T^{\nu\mu}_{\,\,\,\,\alpha} - T^{\alpha}_{\,\,\,\,\mu\alpha} T^{\beta\mu}_{\,\,\,\,\beta}\,,
\end{equation}

which is derived in such a way to be equivalent to the regular curvature-based Ricci scalar (up to a boundary term). This means that TEGR will be defined by an action based on the linear form of $T$.

TEGR can be directly modified to our scalar-tensor form by generalizing it to the action \cite{PhysRevD.97.104011}
\begin{equation}\label{1}
    \mathcal S =\int d^{4}xe[f(T,\phi)+P(\phi)X]+S_{m}\,,
\end{equation}

where $f(T,\phi)$ is an arbitrary function of the torsion scalar $T$ and the scalar field $\phi$, and $X=-\partial_ \mu \phi \partial^ \mu \phi/2$. This broad action includes non-minimally coupled scalar-torsion gravity models with $f(T,\phi)$ coupling function, $f(T)$ gravity, and a minimally coupled scalar field. Here we assume geometric units, and write the tetrad determinant as $e=\text{det}[e^{A}_{\,\,\,\,\mu}]=\sqrt{-g}$

we consider the homogeneous and isotropic flat Friedmann-Lema\^{i}tre-Robertson-Walker (FLRW) geometry in order to proceed to the cosmological application of $f(T,\phi)$ gravity.
\begin{equation}\label{7}
    ds^{2} = -dt^{2}+a^{2}(t)\delta_{\mu \nu} dx^{\mu} dx^{\nu}\,,   
\end{equation}

where $a(t)$ is the scale factor that represents the expansion in the spatial directions. The tetrad, $e^{A}_{\,\,\,\,\mu} = {\rm diag}(1, a(t),\\ a(t), a(t))$. From Eq. \eqref{eq:torsion_scalar_def}, the torsion scalar becomes, $T=6H^2$. Varying the action in Eq. \eqref{1} with respect to the tetrad field and the scalar field $\phi$, the field equations of $f(T,\phi)$ gravity can be obtained along with the Klein-Gordon equation as,

\begin{align}
    f(T,\phi)-P(\phi)X-2Tf_{,T} = \rho_{m}\,, \label{8}\\
    f(T,\phi)+P(\phi)X-2Tf_{,T}-4\dot{H}f_{,T}-4Hf_{,T} = -p_{m}\,, \label{9}\\
    -P_{,\phi}X-3P(\phi)H\dot{\phi}-P(\phi)\ddot{\phi}+f_{,\phi} = 0\,, \label{10}
\end{align}

where  $H = \frac{\dot{a}}{a}$ is the Hubble rate, and an over dot denotes the derivative with respect to cosmic time $t$. A comma indicates the derivative for $T$ or $\phi$. The functions $p_{m}$ and $\rho_{m}$ represent the pressure and energy density of matter respectively. One can refer the Friedmann equations and Klein-Gordon equation of $f(T,\phi)$ gravity in Refs. ~\cite{Gonzalez-Espinoza:2021mwr,Gonzalez-Espinoza:2020jss,Gonzalez_Espinoza_2021ge,Gonzalez_Espinoza_2020}.

\section{Lagrangian formalism of \texorpdfstring{$f(T,\phi)$}{} theory}\label{SEC-III}
The Lagrangian formalism of $f (T,\phi )$ theory has been formulated in this section. The point-like Lagrangian is useful in the analysis of Noether symmetry, which deals with the Friedmann equations, and can be derived from Eq.~(\ref{1}) or followed from Ref.~\cite{Capozziello_2000}. One can establish a Canonical Lagrangian $\mathcal{L}$ = $\mathcal{L}(a, \dot{a}, T, \dot{T},\phi,\dot{\phi})$ to deduce the cosmological equations in the FLRW metric, whereas $\mathcal Q=(a, T,\phi)$ is the configuration space from which it is possible to derive the tangent space denoted by $\mathcal{T}\mathcal{Q}$ and can be obtained as $\mathcal{T}\mathcal{Q}= (a, \dot{a}, T, \dot{T},\phi,\dot{\phi})$, the corresponding tangent space on which $\mathcal{L}$ is defined as an application. Here, the scale factor $a(t)$, torsion scalar $T$, and scalar field $\phi(t)$ are taken as independent dynamical variables of the FLRW metric. One can use the method of Lagrange multipliers to set $T-6\frac{\dot{a}^{2}}{a^{2}}=0$ as a constraint of the dynamics and integrating by parts, the Lagrangian $\mathcal{L}$ becomes analogous to Ref.~\cite{Capozziello_2000}, and so we obtain
\begin{equation}\label{11}
    \mathcal S = 2 \pi^{2} \int a^{3} \left[f(T,\phi)+P(\phi)\frac{\dot{\phi}^{2}}{2}-\lambda \left (T-6\frac{\dot{a}^{2}}{a^{2}} \right )-\frac{\rho_{m0}}{a^{3}} \right ] dt   \,,  
\end{equation}
where $\lambda$ is a Lagrange multiplier and $\rho_{m0}$ is the matter energy density at present time. By varying this action in Eq.~(\ref{11}) with respect to $T$, we get
\begin{equation} \label{12}
    \lambda = f_{,T}(T,\phi)\,.
\end{equation}
Thus, the action in Eq.~(\ref{11}) can be written  as 
\begin{equation}\label{13}
    \mathcal S = 2 \pi^{2} \int a^{3} \left[f(T,\phi)+P(\phi)\frac{\dot{\phi}^{2}}{2}-f_{,T} \left (T-6\frac{\dot{a}^{2}}{a^{2}} \right )-\frac{\rho_{m0}}{a^{3}} \right ] dt \,,
\end{equation}
and the point-like Lagrangian is 
\begin{equation} \label{14}
    \mathcal{L}(a, \dot{a}, T, \dot{T},\phi,\dot{\phi}) = a^{3}\left(f(T,\phi)+P(\phi)\frac{\dot{\phi}^{2}}{2}-Tf_{,T}(T,\phi)\right)+6a \dot{a}^{2} f_{,T}(T,\phi)-\rho_{m0}\,.
\end{equation}
The Euler-Lagrange equation given is 
\begin{equation}\label{15}
    \frac{d}{dt}\left(\frac{\partial \mathcal{L}}{\partial \dot{q_{i}}}\right)-\frac{\partial \mathcal{L}}{\partial q_{i}} = 0\,,
\end{equation}

where $q_{i}$ are the generalized coordinates of the configuration space $\mathcal Q$, and here we consider $q_{i}$= $a$, $T$ and $\phi$. In this case, the equations of motion can be described as, 
\begin{align}
    \frac{d}{dt}\left(\frac{\partial \mathcal{L}}{\partial \dot{a}}\right)-\frac{\partial \mathcal{L}}{\partial a}&=0\,, \label{16} \\
    \frac{d}{dt}\left(\frac{\partial \mathcal{L}}{\partial \dot{T}}\right)-\frac{\partial \mathcal{L}}{\partial T}&=0\,, \label{17} \\
    \frac{d}{dt}\left(\frac{\partial \mathcal{L}}{\partial \dot{\phi}}\right)-\frac{\partial \mathcal{L}}{\partial \phi}&=0\,. \label{18}
\end{align}
Substituting Eq.~(\ref{14}) into Eqs.~(\ref{16}-\ref{18}), we will get Euler-Lagrange equations for $a$, $T$ and $\phi$ as 
\begin{align}
    f(T,\phi)+P(\phi)X-2Tf_{,T}-4\dot{H}f_{,T}-4Hf_{,T}&=0\,, \label{19}\\
    a^{3}f_{,TT} \left(T-6\frac{\dot{a}^2}{a^{2}}\right)&=0\,, \label{20}\\
    -P_{,\phi} \frac{\dot{\phi}^{2}}{2}-3P(\phi)H\dot{\phi}-P(\phi)\ddot{\phi}+f_{,\phi}&=0\,. \label{21}
\end{align}

From Eq.~(\ref{20}), if $f_{,TT} \neq 0$, then we get $T=6\frac{\dot{a}^2}{a^{2}}=6 H^{2}$ which is the torsion scalar of the FLRW metric. On the other hand, from Eqs.~(\ref{19},\ref{21}), we can say that these two relations are the same as Eqs.~(\ref{9}--\ref{10}), i.e., the modified Friedmann equation is recovered with the help of Lagrangian $\mathcal{L}$. The energy conditions of Lagrangian $\mathcal{L}$ are defined by
\begin{equation} \label{22}
    E_{\mathcal{L}}=  \frac{\partial \mathcal{L}}{\partial \dot{a}} \dot{a}+\frac{\partial \mathcal{L}}{\partial \dot{T}} \dot{T}+\frac{\partial \mathcal{L}}{\partial \dot{\phi}} \dot{\phi}-\mathcal{L}\,.
\end{equation}
Now, substituting Eq.~(\ref{14}) into Eq.~(\ref{22}) we find that 
\begin{equation}\label{23}
    E_{\mathcal{L}}(a,\dot{a},T,\dot{T},\phi,\dot{\phi})=12 H^{2} f_{,T}+P(\phi)\frac{\dot{\phi}^{2}}{2}-f(T,\phi)+\frac{\rho_{m0}}{a^{3}}   \,.
\end{equation}
Considering the total energy $E_{\mathcal{L}}=0$, we obtain
\begin{equation}\label{24}
f(T,\phi)-12 H^{2} f_{,T}-P(\phi)\frac{\dot{\phi}^{2}}{2}=\frac{\rho_{m0}}{a^{3}} \,.     
\end{equation}

Here, Eq.~(\ref{24}) is equivalent to Eq.~(\ref{8}), so we conclude that the point-like Lagrange in Eq.~(\ref{14}) can derive all the cosmological equations.

\section{Noether symmetries} \label{SEC-IV}
Noether symmetries often play a significant role in physics since they may be utilized to determine the integrability of a differential equations system and simplify it. Generally, a conserved quantity with a physical meaning can be linked to a Noether symmetry existence. In cosmology, the so-called Noether Symmetry approach is very helpful in finding out exact solutions. We briefly discuss how a general differential equation functions when a point transformation is at work. Consider a system with $n$ generalized coordinates $x^{i}$ and an independent variable $t$ driven by a Lagrangian $\mathcal{L}$. The following is the general form of an infinitesimal change affecting that system. Let us say that the expression for a one-parameter point transformation is
\begin{equation}\label{25}
    \bar{t}= \psi(t,x^{k},\epsilon)\,,  \hspace{1cm} \bar{x}^{A}=\eta(t,x^{k},\epsilon)\,.
\end{equation}
In this scenario, the one-parameter point transformation generating vector is given by
\begin{equation}\label{26}
    \mathcal{Y} = \xi(t,x^{k},\epsilon)\frac{\partial}{\partial t}+\alpha_{i}(t,x^{k},\epsilon)\frac{\partial}{\partial x^{k} }\,,
\end{equation}
where 
\begin{equation}\label{27}
    \xi(t,x^{k})=\frac{\partial \psi(t,x^{k},\epsilon) }{\partial \epsilon} |{\epsilon \to 0}  \hspace{1cm} \alpha_{i}(t,x^{k})=\frac{\partial \eta(t,x^{k},\epsilon) }{\partial \epsilon} |{\epsilon \to 0}\,.
\end{equation}
In this case, the $n^{th}$ prolongation of the generator vector is \cite{Dialektopoulos:2021ryi} 
\begin{equation}\label{28}
    \mathcal{Y}^{[n]}= \mathcal{Y}+\alpha_{i}^{[1]}\frac{\partial}{\partial \dot{x}^{i}}+.....+\alpha_{i}^{[n]}\frac{\partial}{\partial x^{(n)i}}\,,
\end{equation}
where 
\begin{align}
    \alpha_i^{[1]}=\frac{d}{dt}\alpha_{i}-\dot{x}^{i}\frac{d}{dt}\xi\,, \label{29} \\
    \alpha_i^{[n]}=\frac{d}{dt}\alpha^{[n-1]}_{i}-x^{(n)i}\frac{d}{dt}\xi\,, \label{30} 
\end{align}

where $\mathcal{Y}^{[n]}$ is called the $n^{th}$ prolongation of the generator vector ~(\ref{26}). Let Eq.~(\ref{26}) be the generator of an infinitesimal transformation and $\mathcal{L}$=$\mathcal{L}(t,x^{i},\dot{x}^{i})$  be a Lagrangian of a dynamical system. Then the Euler-Lagrange equations are invariant under the transformation if and only if  there exists a function $g = g(t, x^{i})$ such that the following (Rund-Trautman identity) condition holds
\begin{equation}\label{31}
    \mathcal{Y}^{[1]}\mathcal{L}+\mathcal{L}\frac{d\xi (t, x^i)}{dt}=\frac{dg(t, x^i)}{dt}\,,
\end{equation}

here $\mathcal{Y}^{[1]}$ is the first prolongation of Eq.~(\ref{31}). If the generator of Eq.~(\ref{28}) satisfies Eq.~(\ref{31}), then the generator vector represented in Eq.~(\ref{26}) is a Noether symmetry of the dynamical system described by the Lagrangian $\mathcal{L}$. According to the well-known Noether theorem, there will be a constant of motion (Noether charge), namely
\begin{equation}\label{32}
    Q_{0}=  \sum_{i}  \alpha_{i} \frac{\partial \mathcal{L}}{\partial \dot{q}_{i}}=constant\,,
\end{equation}
where $q_{i}$ defined the coordinates of configuration space and $\alpha_{i}$ describes the Noether factors.

\section{Noether Symmetries in \texorpdfstring{$f(T,\phi)$}{} gravity } \label{SEC-V}
This section discusses Noether symmetry in scalar-torsion $f(T,\phi)$ theory. The Noether symmetry technique \cite{Capozziello_1993} is used to determine possible symmetries for the Lagrangian dynamical system ~(\ref{14}). Noether symmetry is a helpful technique for finding the exact solution to a given Lagrangian, and the finding models are justified at a fundamental level. The generator of the Noether symmetry is a vector $\mathcal{Y}$. The presence of symmetry is based on a vector specified on the Lagrangian $\mathcal{L}$ tangent space. This section will examine one-parameter point transformation in the configuration space $(t, a, T,\phi)$. The generator is written as
\begin{equation}\label{33}
    \mathcal{Y}=\xi(t,a,T,\phi)\frac{\partial}{\partial t}+\alpha_{1}(t,a,T,\phi)\frac{\partial}{\partial a}+\alpha_{2}(t,a,T,\phi)\frac{\partial}{\partial T}+\alpha_{3}(t,a,T,\phi)\frac{\partial}{\partial \phi}\,, 
\end{equation}
and the first prolongation of the generator vector is 
\begin{equation}\label{34}
    \mathcal{Y}^{[1]}=\mathcal{Y}+\alpha^{[1]}_{1} \frac{\partial}{\partial \dot{a}}+\alpha^{[1]}_{3} \frac{\partial}{\partial \dot{\phi}}\,,
\end{equation} 
with
\begin{align}\label{35}
    \alpha^{[1]}_{1} = \frac{\partial}{\partial t}\alpha_{1}+\dot{a}\frac{\partial}{\partial a}\alpha_{1}+\dot{\phi}\frac{\partial}{\partial \phi}\alpha_{1}+\dot{T}\frac{\partial}{\partial T} \alpha_{1}-\dot{a}\frac{\partial}{\partial t}\xi-\dot{a}^{2}\frac{\partial}{\partial a}\xi-\dot{a}\dot{\phi}\frac{\partial}{\partial \phi}\xi-\dot{a} \dot{T}\frac{\partial}{\partial T}\xi \,,\\ 
    \alpha^{[1]}_{3} = \frac{\partial}{\partial t}\alpha_{3}+\dot{a}\frac{\partial}{\partial a}\alpha_{3}+\dot{\phi}\frac{\partial}{\partial \phi}\alpha_{3}+\dot{T}\frac{\partial}{\partial T} \alpha_{3} -\dot{\phi}\frac{\partial}{\partial t}\xi-\dot{a} \dot{\phi}\frac{\partial}{\partial a}\xi-\dot{\phi}^{2}\frac{\partial}{\partial \phi}\xi-\dot{\phi} \dot{T}\frac{\partial}{\partial T}\xi \label{36}\,. 
\end{align}
We calculate each term in the symmetry condition from Eq. (\ref{31}). The first term $\mathcal{Y}^{[1]}\mathcal{L}$ is
\begin{align} \label{37}
    \mathcal{Y}^{[1]} \mathcal{L} &= 3 a^{2} \alpha_{1}f-3 \alpha_{1} a^{2} T f_{,T}-a^{3} \alpha_{2} T f_{,TT} +\alpha_{3} a^{3} f_{,\phi}-a^{3} \alpha_{3} T f_{,T\phi} \nonumber \\
    & +12 a \dot{a} \dot{T} f_{,T} \frac{\partial \alpha_{1}}{\partial T}-12 a \dot{a}^{3} f_{,T} \frac{\partial \xi}{\partial a}-12 a \dot{a}^{2} \dot{\phi} f_{,T} \frac{\partial \xi }{\partial \phi}-12 a \dot{a}^{2}\dot{T} f_{,T} \frac{\partial \xi}{\partial T}\nonumber \\
    & + \left(\frac{3}{2}\alpha_{1}a^{2}P(\phi)+\frac{1}{2} a^{3}\alpha_{3}P_{,\phi} +a^{3}P(\phi)\frac{\partial \alpha_{3}}{\partial \phi}  -a^{3} P(\phi) \frac{\partial \xi}{\partial t} \right ) \dot{\phi}^{2}\nonumber \\
    & + \left(6 \alpha_{1} f_{,T} +6 \alpha_{2} a f_{,TT}+ 6 \alpha_{3} a f_{,T\phi}+12af_{,T}\frac{\partial \alpha_{1}}{\partial a}-12 a f_{,T} \frac{\partial \xi}{\partial t}  \right) \dot{a}^{2}\nonumber\\
    &   + 12 a \dot{a} f_{,T} \frac{\partial \alpha_{1}}{\partial t} +\left(12 a f_{,T} \frac{\partial \alpha_{1}}{\partial \phi}+ a^{3} P(\phi) \frac{\partial \alpha_{3}}{\partial a}\right)\dot{a} \dot{\phi} +a^{3} P(\phi) \frac{\partial \alpha_{3}}{\partial t}  \dot{\phi}\nonumber \\
    &+a^{3}P(\phi) \dot{\phi} \dot{T} \frac{\partial \alpha_ {3}}{\partial T}-a^{3} P(\phi) \dot{\phi}^{3} \frac{\partial \xi}{\partial \phi}-a^{3} P(\phi) \dot{\phi}^{2} \dot{T} \frac{\partial \xi}{\partial T} - a^{3} P(\phi) \dot{a} \dot{\phi}^{2} \frac{\partial \xi}{\partial a}\,,
\end{align}
and the second term of Eq. (\ref{31}) $\mathcal{L} \dot{\xi}$ is
\begin{align}\label{38}
    \mathcal{L} \dot{\xi}&= \left(a^{3} f-a^{3} T f_{,T} +a^{3} P(\phi) \frac{\dot{\phi}^{2}}{2}+6 a \dot{a}^{2} f_{,T}\right) \frac{\partial \xi}{\partial t}\nonumber \\
    & +\left(a^{3}\dot{a} f-a^{3} \dot{a}T f_{,T} +a^{3}\dot{a} P(\phi) \frac{\dot{\phi}^{2}}{2}+6 a \dot{a}^{3} f_{,T}\right) \frac{\partial \xi}{\partial a}\nonumber \\
    & +\left(a^{3} \dot{T} f-a^{3}\dot{T} T f_{,T} +a^{3}\dot{T} P(\phi) \frac{\dot{\phi}^{2}}{2}+6 a \dot{T}\dot{a}^{2} f_{,T}\right) \frac{\partial \xi}{\partial T}\nonumber \\
    & +\left(a^{3}\dot{\phi} f-a^{3} \dot{\phi}T f_{,T} +a^{3}\dot{\phi} P(\phi) \frac{\dot{\phi}^{2}}{2}+6 \dot{\phi}a \dot{a}^{2} f_{,T}\right) \frac{\partial \xi}{\partial \phi}\,.
\end{align}
Furthermore, the right-side of Eq. (\ref{31}) is
\begin{equation}\label{39}
    \dot{g}= \frac{\partial g}{ \partial t}+\dot{a} \frac{\partial g}{ \partial a}+ \dot{T}\frac{\partial g}{ \partial T}+\dot{\phi} \frac{\partial g}{ \partial \phi} \,. 
\end{equation}
 
We obtain the following set of Noether symmetry conditions by substituting the outcomes in Eq. ~(\ref{31}) and setting the terms with the powers of $\dot{a}^{2}$, $\dot{T^{2}}$, $\dot{\phi}^{2}$, $\dot{a}\dot{T}$, $\dot{a}\dot{\phi}$ and  $\dot{\phi}\dot{T}$ equal to zero in order to  choose the generator vector.
\begin{align}
    12 a f_{,T} \frac{\partial \alpha_{1}}{\partial T}=0\,, \hspace{1cm} 6 a f_{,T} \frac{\partial \xi}{\partial a}=0\,, \hspace{1cm}  6 a f_{,T} \frac{\partial \xi}{\partial \phi}=0\,,  \hspace{1cm} 6 a f_{,T} \frac{\partial \xi}{\partial T}=0\,, \label{40} \\
    a^{3} P(\phi)\frac{\partial \alpha_{3}}{\partial T}=0\,, \hspace{1cm}  a^{3} P(\phi)\frac{\partial \xi}{\partial a}=0\,, \hspace{1cm} a^{3} P(\phi)\frac{\partial \xi}{\partial \phi}=0\,, \hspace{1cm} a^{3} P(\phi)\frac{\partial \xi}{\partial T}=0\,, \label{41} \\
    3 a^{2} \alpha_{1}f-3 \alpha_{1} a^{2} T f_{,T}-a^{3} \alpha_{2} T f_{,TT} +\alpha_{3} a^{3} f_{,\phi}-a^{3} \alpha_{3} T f_{,T\phi}+a^{3} f \frac{\partial \xi}{\partial t}-a^{3} T f_{,T}\frac{\partial \xi}{\partial t} = \frac{\partial g}{\partial t}\,, \label{42}\\
    3\alpha_{1}a^{2}P(\phi)+ a^{3}\alpha_{3}P_{,\phi} +2 a^{3}P(\phi)\frac{\partial \alpha_{3}}{\partial \phi}  -a^{3} P(\phi) \frac{\partial \xi}{\partial t}=0\,, \label{43}\\
    6 \alpha_{1} f_{,T} +6 \alpha_{2} a f_{,TT}+ 6 \alpha_{3} a f_{,T\phi}+12af_{,T}\frac{\partial \alpha_{1}}{\partial a}-6 a f_{,T} \frac{\partial \xi}{\partial t} =0\,, \label{44}\\
    12 a f_{,T} \frac{\partial \alpha_{1}}{\partial t}+a^{3} f \frac{\partial \xi}{\partial a}-a^{3} T f_{,T} \frac{\partial \xi}{\partial a}=\frac{\partial g}{\partial a}\,, \label{45}\\
    12 a f_{,T} \frac{\partial \alpha_{1}}{\partial \phi}+ a^{3} P(\phi) \frac{\partial \alpha_{3}}{\partial a}=0\,, \label{46}\\
    a^{3} P(\phi) \frac{\partial \alpha_{3}}{\partial t}+a^{3} f \frac{\partial \xi}{\partial \phi}-a^{3} T f_{,T} \frac{\partial \xi}{\partial \phi}= \frac{\partial g}{\partial \phi}\,, \label{47}\\
    a^{3}f\frac{\partial \xi}{\partial T}- a^{3} T f_{,T} \frac{\partial \xi}{\partial T}=\frac{\partial g}{\partial T}\,. \label{48}
\end{align}
 
Here, the unknown variables are $\xi(t,a,T,\phi)$, $\alpha_{1}(t,a,T,\phi)$, $\alpha_{2}(t,a,T,\phi)$, $\alpha_{3}(t,a,T,\phi)$  and the function  $f(T, \phi)$. If at least one of the variables is non-zero, then we can say that Noether symmetry exists. There are two methods for figuring it out and discovering symmetry. First, the system of partial differential equations [\ref{40}-\ref{48}] may be solved directly, and then the unknown variables and functions can be obtained. In a second strategy, imposing particular forms of $f(T, \phi)$ and discovering related symmetries is possible. From equations (\ref{40}) and (\ref{41}), we can say that Noether coefficients $\alpha_{1}$ and $\alpha_{3}$ are independent to torsion scalar $T$ and also $\xi$ is independent to $a$, $T$ and $\phi$. That means $\alpha_{1}=\alpha_{1}(t,a,\phi)$, $\alpha_{3}=\alpha_{3}(t,a,\phi)$ and $\xi=\xi(t)$. In this work, we will adopt a second strategy to discuss the symmetries in $f(T, \phi)$ cosmology. To achieve this, we will consider the specific forms of $f(T,\phi)$ and find the solution of the system of partial differential equations  [\ref{40}-\ref{48}]. Here, we are also considering $P(\phi)=1$ to solve a system of partial differential equations.  

\section{Cosmological model $f(T,\phi)=-T F(\phi) +V(\phi)$} \label{SEC-VI}
In the above form of $f(T,\phi)$, $F(\phi)$ is the non-minimal coupling function of scalar filed $\phi$ and $V(\phi)$ are the scalar potential functions. In this study, we have taken $F(\phi)=f_{0} \phi^{2}$ and $V(\phi)=V_{0} \phi^{m}$, then we can rewrite $f(T,\phi)=-T f_{0} \phi^{2} +V_{0} \phi^{m}$, where $f_{0}$, $V_{0}$ and $m$ are arbitrary constants. This $f(T,\phi)$ choice comes from Ref. \cite{PhysRevD.97.104011, Gonzalez_Espinoza_2020}. We insert this form of $f(T,\phi)$ in the system of partial differential equations (\ref{40}-\ref{48}), and using the separation of the variable method, we get the Noether coefficients of the Noether vector (\ref{33}). In this case, the function $g$ remains constant and is defined by $g_0$, and $\beta_{0}$ is an integration constant.
\begin{align}\label{49}
     \alpha_{1}(t,a,T,\phi)&=\beta_{0} \frac{a}{3}\,,  \\ \nonumber
     \alpha_{2}(t,a,T,\phi)&=\alpha_{2}(t,a,T,\phi)\,,   \\ \nonumber 
     \alpha_{3}(t,a,T,\phi)&= -\frac{ \beta_{0} \phi}{m}\,,  \\ \nonumber
     \xi(t,a,T,\phi)&= \beta_{0} t\,, \\ \nonumber
     g(t,a,T,\phi)&= g_{0}\,, \\ \nonumber 
\end{align}

From Eq.\eqref{49}, we have obtained the condition on the model parameter $m\neq0$. These Noether coefficients substituted in Eq.~(\ref{33}) to obtain the Noether vector as,
\begin{equation}\label{50}
     \mathcal{Y}=\beta_{0}t\frac{\partial}{\partial t}+\beta_{0} \frac{a}{3}\frac{\partial}{\partial a}+\alpha_{2}\frac{\partial}{\partial T}-\frac{\beta_{0} \phi}{m}\frac{\partial}{\partial \phi}\,.
\end{equation}

 Let us now look for a cosmological solution to this kind of function. The point-like Lagrangian in Eq.~(\ref{14}) looks like
 \begin{equation}\label{51} 
     \mathcal{L} = a^{3} V_{0} \phi^{m}+a^{3}\frac{\dot{\phi}^{2}}{2}-6 a \dot{a}^{2} f_{0} \phi^{2}-\rho_{m0}     \,.
\end{equation}
The Euler-Lagrange equation for the scale factor Eq.~(\ref{19}) and energy density Eq.~(\ref{24}) are given as 
\begin{align}
    2 f_{0} \phi^{2}\frac{\dot{a}^{2}}{a^{2}}+4 f_{0} \phi^{2} \frac{\ddot{a}}{a}+4 f_{0} \phi^{2} \frac{\dot{a}}{a}+V_{0}\phi^{m}+\frac{\dot{\phi}^{2}}{2}=0\,,\label{52} \\
     6 f_{0} \phi^{2}\frac{\dot{a}^{2}}{a^{2}}+V_{0}\phi^{m}-\frac{\dot{\phi}^{2}}{2}-\frac{\rho_{m0}}{a^{3}}=0\,. \label{53}
 \end{align}
 
Finding a solution to the dynamical equations (\ref{52}-\ref{53}) is challenging since they are non-linear differential equations. To solve this problem, we need more variables in Lagrangian (\ref{14}). When the Noether symmetry exists, we can use a cyclic variable to change the coordinates. Following Ref. \cite{PhysRevD.42.1091, Kucukakca_2013}, we perform the coordinate transformation $(a, \phi)$ $\to$ $(u, v)$, where $u$ is a cyclic variable. The partial differential equations generated by such a transformation are as follows:
\begin{align}
\alpha_{1} \frac{\partial u}{\partial a}+\alpha_{3} \frac{\partial u}{\partial \phi}=1\,,\label{54} \\
\alpha_{1} \frac{\partial v}{\partial a}+\alpha_{3} \frac{\partial v}{\partial \phi}=0\,, \label{55} 
\end{align}

where the new variables $u$ and $v$ are functions of old variables $a$ and $\phi$. we have  obtained the solution of Eqs. (\ref{54}-\ref{55}) as
\begin{equation}
u(a,\phi)= \frac{3 \ln(a)}{\beta_{0}}, \hspace{1cm} v(a,\phi)=\phi a^{\frac{3}{m}} \,, \label{56}
\end{equation}

when $a$ and $\phi$ are transformed into the new variables $u$ and $v$, 
\begin{equation}
a(u,v)= e^{\frac{\beta_{0} u}{3}}, \hspace{1cm} \phi(u,v)=v e^{-\frac{u \beta_{0}}{m}} \,, \label{57}
\end{equation}

It is important to remember that when the transformation mentioned above is used, the variable $u$ does not show up in the Lagrangian \eqref{14} because, in this study, we consider $u$ as a cyclic variable. From this point onward, we will take the model parameter $f_{0}=\frac{3}{16}$, and we will also discuss the special case $m=2$. This transformation makes it possible to write the Lagrangian \eqref{51} in the format shown below:
\begin{equation}
\mathcal{L}=V_{0}v^{2}+\frac{1}{2}\dot{v}^{2}-\frac{1}{2} v \dot{v} \dot{u} \beta_{0}-\rho_{m0}\,, \label{58}   
\end{equation}

It is clear from this Lagrangian that it is independent of the cyclic variable $u$. From this Lagrangian, we have obtained the corresponding field equations are
\begin{align}
v \dot{v} \beta_{0}=-2 Q_{0}\,,\label{59}\\ 
\ddot{v}-\frac{1}{2} v \ddot{u} \beta_{0}-2V_{0}v=0\,, \label{60}\\
\dot{v}v\dot{u}\beta_{0}-\dot{v}^{2}+2 V_{0}v^{2}-2 \rho_{m0}=0\,,\label{61}
\end{align}

where $Q_{0}$ is a constant corresponding to a conservative quantity. From Eq. (\ref{59}), we get
\begin{equation}\label{62}
v(t)=\left(\frac{-4 Q_{0} t}{\beta_{0}}+2 v_{1}\right)^{\frac{1}{2}}\,,    
\end{equation}

where $v_{1}$ is an integration constant. We determine $u(t)$ by inserting the solution \eqref{62} into Eq. \eqref{60}
\begin{equation}\label{63}
u(t)=-\frac{2 \left(-\frac{1}{4} \ln \left(4 Q_0 t-2 \beta_{0} v_1\right)-\frac{\beta_{0} t v_1 V_0}{Q_0}+t^2 V_0\right)}{\beta_{0}}+u_0 t+u_1\,,    
\end{equation}

where $u_{0}$ and $u_{1}$ are an integration constant. From Eq. \eqref{61}, we have the following restriction, 
\begin{equation}\label{64}
-2 \left(\rho _{\text{m0}}+u_0 Q_0\right)=0\,,    
\end{equation}

It is evident from this restriction that $u_{0}$ must be zero in the absence of the standard matter. By substituting the solution $v(t)$ and $u(t)$, which is presented in Eqs. (\ref{62}-\ref{63}) into Eq. \eqref{57}, we obtain the cosmological solution as,
\begin{eqnarray}
a(t)&=& e^{\frac{1}{3} \left(\beta_{0} u_1+\frac{\ln \left(4 Q_0 t-2 \beta_{0} v_1\right)-4 t^2 V_0}{2 \beta_{0}}+\frac{2 t v_1 V_0}{Q_0}+u_0 t\right)}\,,\label{65}   \\
\phi(t)&=& \sqrt{2} \sqrt{v_1-\frac{2 Q_0 t}{\beta_{0}}} e^{-\frac{1}{4} \ln \left(4 Q_0 t-2 \beta_{0} v_1\right)+t V_0 \left(t-\frac{\beta_{0} v_1}{Q_0}\right)-\frac{1}{2} \beta_{0} (u_0 t+u_1)}\,. \label{66}   
\end{eqnarray}

The aforementioned solutions contain the three integration constants $v_{1}$, $u_{0}$, and $u_{1}$. To determine the integration constant in the general solution of the filed equations (\ref{65}-\ref{66}) by following the steps in Ref. \cite{Rubano_2004}. We start by considering the scenario when $a(0) = 0$, which fixes the origin of time. It is best to consider this condition a random selection at the beginning of time. This condition is applied to the scale factor \eqref{65} yields $v_{1}=0$. Next, we set the present time $t_{0}=1$. Thus, we may assume that $a(t_{0} = 1) = 1$ is the norm. An expression that results from this condition is
\begin{equation}\label{67}
u_{1}=\frac{4 V_{0}-2 u_{0} \beta_{0}-\ln(4 Q_{0})}{2 \beta_{0}^{2}}\,.    
\end{equation}

The last condition is to set $H(t_{0} = 1)=\mathcal{H}_{0} $, where $H(t)$ is the Hubble parameter. Here, the parameter $\mathcal{H}_{0}$ is not the same as the standard observations of the Hubble constant $\mathcal{H}_{0}$. This condition applies in Eq. \eqref{65}, then we have obtained
\begin{equation} \label{68}
u_{0}=\frac{6 \beta_{0}\mathcal{H}_{0}-1-8 V_{0}}{2 \beta_{0}}\,.    
\end{equation}

When these constraints are applied, the scale factor \eqref{65} and the scalar field \eqref{66} can be written as follows:
\begin{eqnarray}
a(t)&=& (4t)^{\frac{1}{6 \beta_{0}}} Q_{0}^{\frac{-1}{2 \beta_{0}}} e^{(t-1)[1+4 V_{0}(3+t)-6\mathcal{H}_{0}\beta_{0}]}\,, \label{69}\\
\phi(t)&=& 2 \sqrt{-\frac{Q_0 t}{\beta_{0}}} e^{\frac{\beta_{0} \left(\mathcal{H}_{0} \left(6-6 \beta_{0} t\right)-\log \left(4 Q_0 t\right)+t\right)+4 Q_0 \ln +4 V_0 \left(\beta_{0} t (t+2)-3\right)-1}{4 \beta_{0}}}\,. \label{70}
\end{eqnarray}

The exact solutions in Equations \eqref{69} and \eqref{70} can be used to 
create all the physical quantities such as the Hubble parameter, deceleration
parameter, and effective equation of state parameter
\begin{eqnarray}
H(t)&=&-6 \beta_{0} \mathcal{H}_{0}+\frac{1}{6 \beta_{0} t}+8 (t+1) V_0+1\,, \label{71}\\
q(t)&=&-1+ \frac{6 \beta_{0} \left(1-48 \beta_{0} t^2 V_0\right)}{\left(6 \beta_{0} t \left(-6 \beta_{0} \mathcal{H}_{0}+8 (t+1) V_0+1\right)+1\right){}^2}\,, \label{72}\\
\omega_{eff}(t)&=&-1+\frac{4 \beta_{0} \left(1-48 \beta_{0} t^2 V_0\right)}{\left(6 \beta_{0} t \left(-6 \beta_{0} \mathcal{H}_{0}+8 (t+1) V_0+1\right)+1\right){}^2}\,. \label{73}
\end{eqnarray}

\begin{figure}[H]
\centering
\includegraphics[width=85mm]{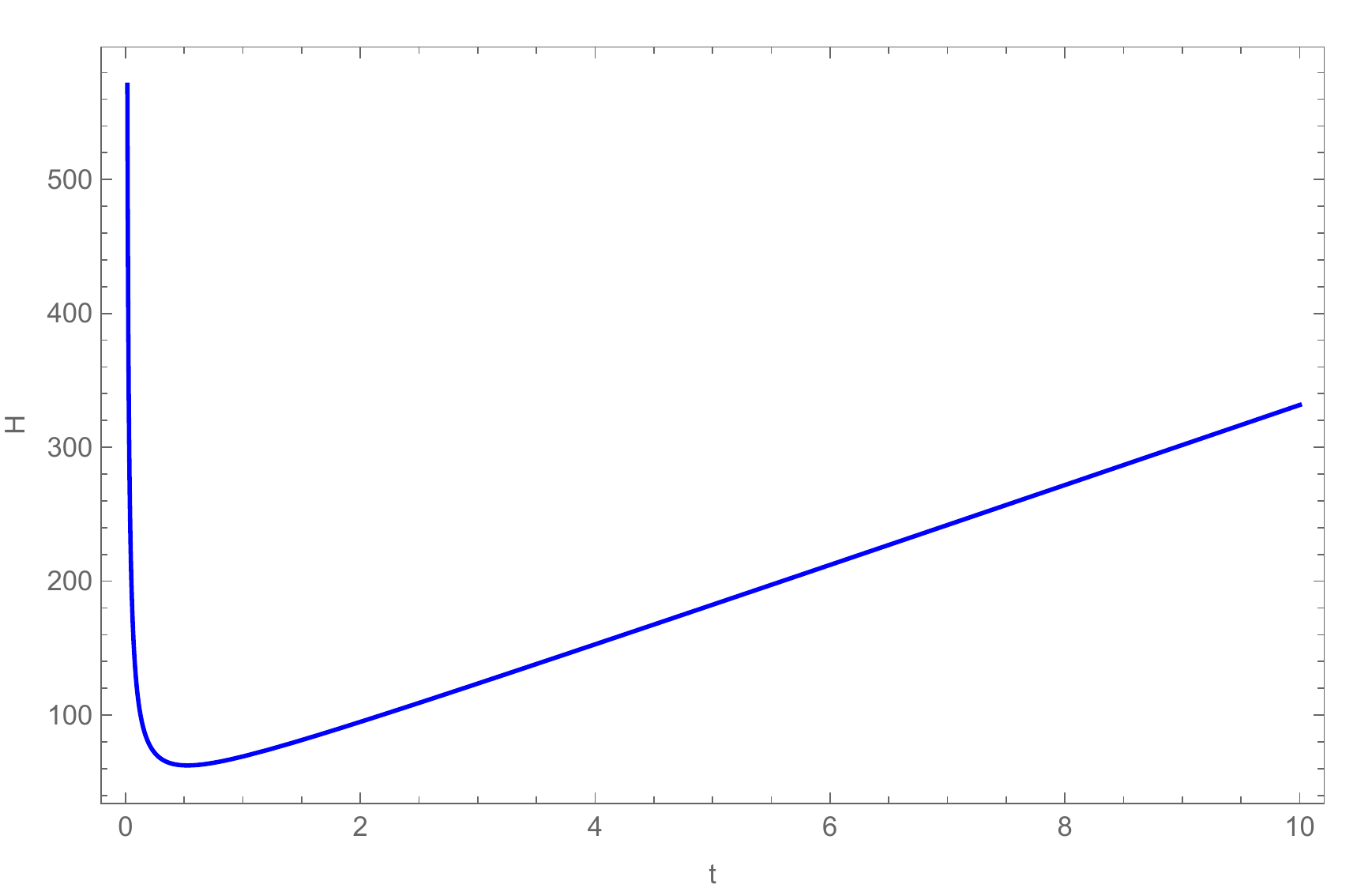}
\includegraphics[width=85mm]{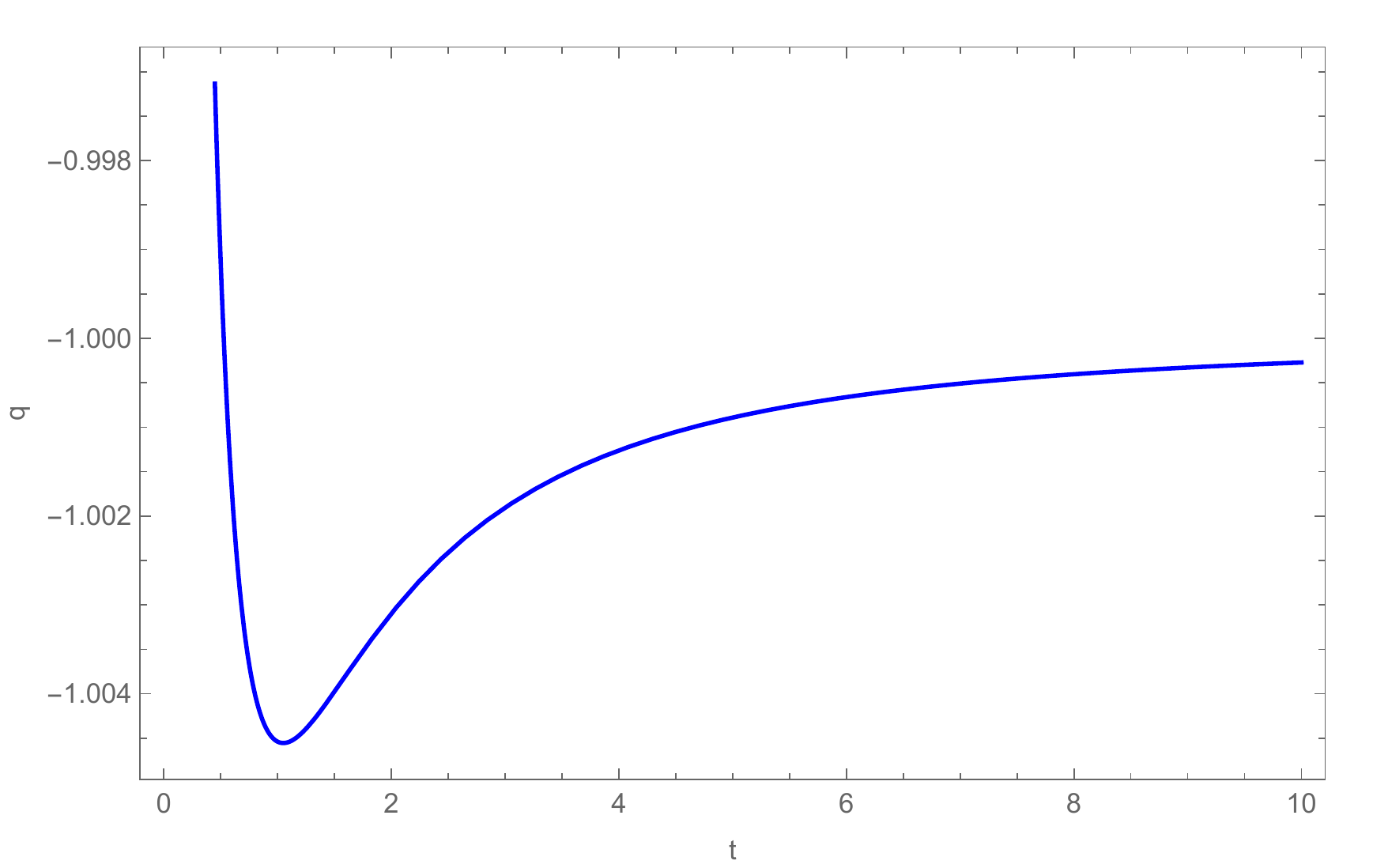}
\caption{Hubble parameter (left panel) and deceleration parameter (right panel) with cosmic time $t$. The parameter scheme: $\beta_{0}=0.02$, $V_{0}=3.75$, and $\mathcal{H}_{0}=2$. } \label{FIG1}
\end{figure}

\begin{figure}[H]
\centering
\includegraphics[width=85mm]{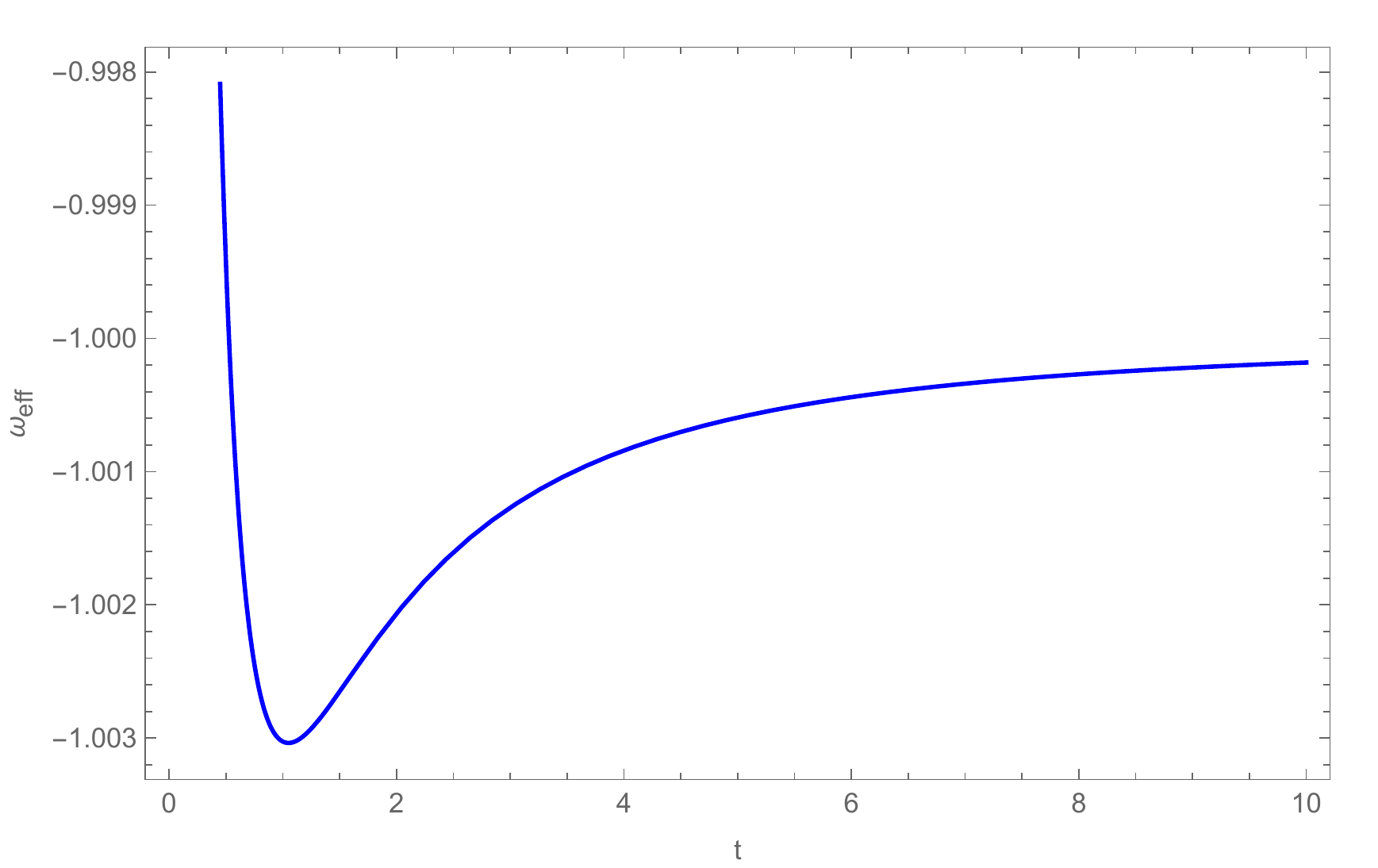}
\caption{Effective equation of state parameter with cosmic time $t$. The parameter scheme: $\beta_{0}=0.02$, $V_{0}=3.75$, and $\mathcal{H}_{0}=2$. } \label{FIG2}
\end{figure} 

The graphical behavior of the physical parameters are given in Fig. \ref{FIG1} and Fig. \ref{FIG2}. Fig.\ref{FIG1} (left panel) shows that the Hubble parameter increases over time, and the present value of the Hubble parameter at $t_{0}=1$ is $69.67$. At $t=0$, the Hubble parameter described the behaviour at early stages of the Universe. During the early stage, when $t$ approaches to zero, the scale factor $a(t)$ is extremely small. Also when $a(t)$ approaches to zero, the Hubble parameter tends towards infinity, indicating an infinite expansion rate at $t=0$. So, from Fig. \ref{FIG1} (left panel), we have observed that the Hubble parameter goes to infinity at cosmic time $t=0$. The Hubble parameter describes the physical explanation for the divergence of the expansion rate around cosmic time $t=0$. It relates to the concept of the Big Bang at the early Universe. In future, the Hubble parameter increasing further, which indicates that the expansion rate $\dot{a}(t)$ is greater than $a(t)$. From the cosmological observations, we can say  that dark energy continues to behave similarly to its current description ( i,e like a cosmological constant or something similar). Dark energy is believed to be the dominant component of the Universe, responsible for its accelerated expansion. This would lead to an exponential expansion of the Universe, where the expansion becomes so rapid that it eventually may tear apart all bound structures and may lead to the rip cosmology. The negative phase of the deceleration parameter in Fig.\ref{FIG1} (right panel) indicates the accelerating era of the Universe, and we have obtained the present value of the deceleration parameter as $-1.0045$. From Fig. \ref{FIG2}, it is evident that the effective equation of state parameter shows a transition behavior from the quintessence phase, ( $\omega_{eff}=>-1$), to the phantom phase, ($\omega_{eff}=<-1$). The present value of the effective equation of the state parameter is noted as $-1.003$. It has been observed that the value of the geometrical parameters obtained at the current time is within the range of the cosmological observations \cite{Aghanim:2018eyx, PhysRevResearch.2.013028, Denzel_2020}.

\section{Conclusion} \label{SEC-VII}
This paper studied the Noether symmetries approach in scalar-torsion $f(T,\phi)$ gravity. The Noether symmetry is one of the most effective mathematical strategies for detecting conserved quantities and simplifying dynamical systems. The Noether symmetry approach is a useful tool to classify the models and find the exact cosmological solution of the field equations. The Lagrangian plays an important role in describing symmetries and the Noether vector in the Noether symmetry. This way, Lagrangian multipliers significantly transform the Lagrangian into Canonical form, as explored in Sec.~\ref{SEC-III}. 

In this work, we explore $f(T,\phi)$ theory, which allows for non-minimal coupling between the torsion scalar and the scalar field. The torsion scalar defines the TEGR action, which effectively means that we allow the scalar field to be dynamic in some instances in the evolution of the Universe. We develop the Lagrangian for the FLRW space-time metric described in Eq.~(\ref{14}). In Sec.~\ref{SEC-V}, we examine the Rund-Trautman identity in Eq.~(\ref{31}) to obtain the system of partial differential Eqs.~(\ref{40}--\ref{48}) that defines the governing equations of the system. In this system of partial differential equations, the unknown variables are $\xi$, $\alpha_{1}$, $\alpha_{2}$, $\alpha_{3}$  and the function $f(T, \phi)$. These variables are Noether coefficients, and $f(T,\phi)$ is an arbitrary function of the torsion scalar and scalar field.

In Sec.~\ref{SEC-VI}, we have discussed the Noether symmetry in $f(T,\phi)$ gravity for a specific form of $f(T,\phi)=-T f_{0} \phi^{2} +V_{0} \phi^{m}$. For this choice of $f(T,\phi)$, we have obtained a nontrivial Noether vector described in Eq.~(\ref{50}). To simplify the dynamical equations of the model under consideration, we may also create a coordinate transformation, as shown in Eqs. \eqref{56} according to the Noether symmetry. A cyclic coordinate is one of the new coordinates. Using these transformations, we could make a new set of field equations (\ref{59}-\ref{61}) for the FLRW metric and find analytical solutions for the scale factor and the scalar field by adjusting the values of $m=2$ and $f_{0}=\frac{3}{16}$, which is described in  Eq.~(\ref{65}- \ref{66}). To learn more about the evolution of the Universe, we also looked at various cosmological parameters. In Fig.\ref{FIG1} and Fig. \ref{FIG2}, we have shown how certain significant cosmological parameters have changed through cosmic time. From Fig. \ref{FIG2}, we can say that the equation of state parameter shows the transition from the quintessence phase to the phantom phase, and we have obtained $\omega_{eff}(t_{0})=-1.003$. In Fig.\ref{FIG1} right panel, we observe that the deceleration parameter shows the accelerating phase of the Universe. The deceleration and Hubble parameter corresponding present values are $q(t_{0})=-1.0045$ and $H(t_{0})=69.67$. These values of the geometrical parameters obtained at the present time ($t_{0}=1$) have been shown to fit inside the range of cosmological observations. The Noether symmetry technique is also used to evaluate alternative physically possible $f(T,\phi)$ forms, which may simplify the dynamics of the system and be useful for understanding the cosmological solution in this scalar-torsion theory.

\section*{Acknowledgements}
LKD acknowledges the financial support provided by University Grants Commission (UGC) through Junior Research Fellowship UGC Ref. No.: 191620180688 to carry out the research work. BM acknowledges the support of IUCAA, Pune (India) through the visiting associateship program. This paper is based upon work from COST Action CA21136 {\it Addressing observational tensions in cosmology with systematics and fundamental physics} (CosmoVerse) supported by COST (European Cooperation in Science and Technology). The authors are thankful to the anonymous referees for their valuable comments and suggestions for the improvement of the paper. 

\bibliographystyle{utphys}
\bibliography{references}
\end{document}